# Impact of room size on WDM optical wireless links with multiple access points and angle diversity receivers

Osama Zwaid Alsulami[1], Mansourah K. A. Aljohani[1], Sarah O. M. Saeed[1], Sanaa Hamid Mohamed[1], T. E. H. El-Gorashi[1], Mohammed T. Alresheedi[2] and Jaafar M. H. Elmirghani[1]
[1]School of Electronic and Electrical Engineering, University of Leeds, LS2 9JT, United Kingdom
[2]Department of Electrical Engineering, King Saud University, Riyadh, Kingdom of Saudi Arabia
**ml15ozma@leeds.ac.uk, ml16mka@leeds.ac.uk, elsoms@leeds.ac.uk, elshm@leeds.ac.uk, t.e.h.elgorashi@leeds.ac.uk, malresheedi@ksu.edu.sa, j.m.h.elmirghani@leeds.ac.uk**

## ABSTRACT
Optical wireless communication (OWC) systems have been the subject of attention as a promising wireless communication technology that can offer high data rates and support multiple users simultaneously. In this paper, the impact of room size is investigated when using wavelength division multiple access (WDMA) in conjunction with an angle diversity receiver (ADR). Four wavelengths (red, yellow, green and blue) can be provided in this work based on the RYGB LDs transmitter used. Three room sizes are considered with two 8-user scenarios. A mixed-integer linear programming (MILP) model is proposed for the purpose of optimising the resource allocation. The optical channel bandwidth, SINR and data rate have been calculated for each user in both scenarios in all rooms. Room A, which is the largest room, can provide a higher channel bandwidth and SINR compared to the other rooms. However, all rooms can provide a data rate above 5 Gbps in both scenarios.

## 1. INTRODUCTION

In recent years, researchers and industry have shown an increased interest in optical wireless communication (OWC), due to the immense growing demand for higher data rates in wireless communication networks. This has led to the need for carriers that are more efficient than the radio frequency (RF) to solve one of the main RF drawbacks which is the channel capacity limitation due to the lack of radio spectrum. Some of the major communication companies and providers such as Cisco have expected the demand for mobile internet traffic to rapidly increase by up to 27 times more than what was used in 2016 [1]. Due to this, researchers have been working on expanding the frequency range available for communication beyond the conventional radio waves

OWC systems allow access to license-free bandwidth and have high levels of security and are relatively low cost compared to conventional radio frequency systems [2]–[9]. Despite the benefits of using OWC systems, some limitations remain, such as the availability and visibility of the line-of-sight (LOS) components in the link which can greatly impact system performance and the transmission of traffic [11], [12]. Furthermore, OWC systems can experience inter-symbol interference (ISI) due to the existence of multi-path propagation. One of the major benefits of using visible light communication (VLC) systems, a type of OWC systems, is the ability to transmit video, data and voice, at data rates up to 25 Gbps in indoor environments, as stated in [8], [9], [18], [19], [10] – [17]. Improved link budgets can be achieved through different forms of adaptation such as beam power, beam angle and beam delay adaptation [20] – [22]. Further improvements in the link budget can be obtained through the use of diversity techniques such as receiver angle diversity [23], [24] which can be used to select the best directions to use to receive the signal and reject interference. The design of the uplink was considered in [25], however more attention has to be given to energy efficiency [26]. Multi-carrier code division multiple access (MC-CDMA) has been considered to enable multi-user optical wireless communication [27], [28]. When considering multiple users in a VLC system, interference between users can degrade the system performance. As a result, several techniques have been investigated in many works to support multiple users and reduce interferences [5], [29] – [33]. Wavelength division multiple access (WDMA) is one solutions that can help in this direction. Also, using a technique that can optimise the resource allocation in conjunction with WDMA can improve the system performance by reducing the interference between users.

In this paper, the impact of room size on WDM optical wireless links with multiple access points and angle diversity receivers is investigated. Red, yellow, green, and blue (RYGB) Laser Diodes (LDs) are the key elements of the optical transmitter used, which provides white colour for indoor illumination and high bandwidth modulation for communication [34]. An angle diversity receiver (ADR) is the examined optical receiver in this work. A WDMA scheme is used to provide multiple access and a Mixed Integer Linear Programme (MILP) is used to optimise the resource allocation. The rest of this paper is organised as follows: Section 2 describes the system configuration. Section 3 shows the simulation results and the conclusions are provided in Section 4.

## 2. SYSTEM CONFIGURATION

In this paper, we considered three different room sizes for the analysis, all rooms are unfurnished and (Room A), (Room B) and (Room C) do not have doors and windows. As can be seen from Figures 1 (a), (b) and (c), respectively, the height of all three rooms is 3 m while the areas of the rooms are (8m x 4m), (4m x 4m) and (8m x 2m) (width x length) respectively. Ray tracing was used for modelling the optical indoor channel similar to [34] and [35]. In this work, reflections up to the second order were considered as reflections higher than the second order have only a little impact on the received optical power [35]. Each surface of each room was divided into small equal elements that reflect light rays similar to plaster walls which reflect light rays closer to the Lambertian pattern as shown in [36]. Thus, these small elements act as small emitters by reflecting the light rays. The elements area can affect the results' resolution. When the area of the elements decreases, the resolution of the results increases. However, by reducing the area of the elements, the simulation computation time increases. The simulation parameters are shown in Table 1 for all rooms. All the communication in the three different rooms is done above the communication floor (CF) as shown in Figures 1a, b and c. An angle diversity receiver (ADR) with 4 branches of narrow field of view (FOV) photodetectors (See Figure 1d) is used. Each photodetector is oriented at a different direction to cover a different transmitter in the ceiling. The optical receiver and transmitter (access point (AP)) parameters are illustrated in Table 1.

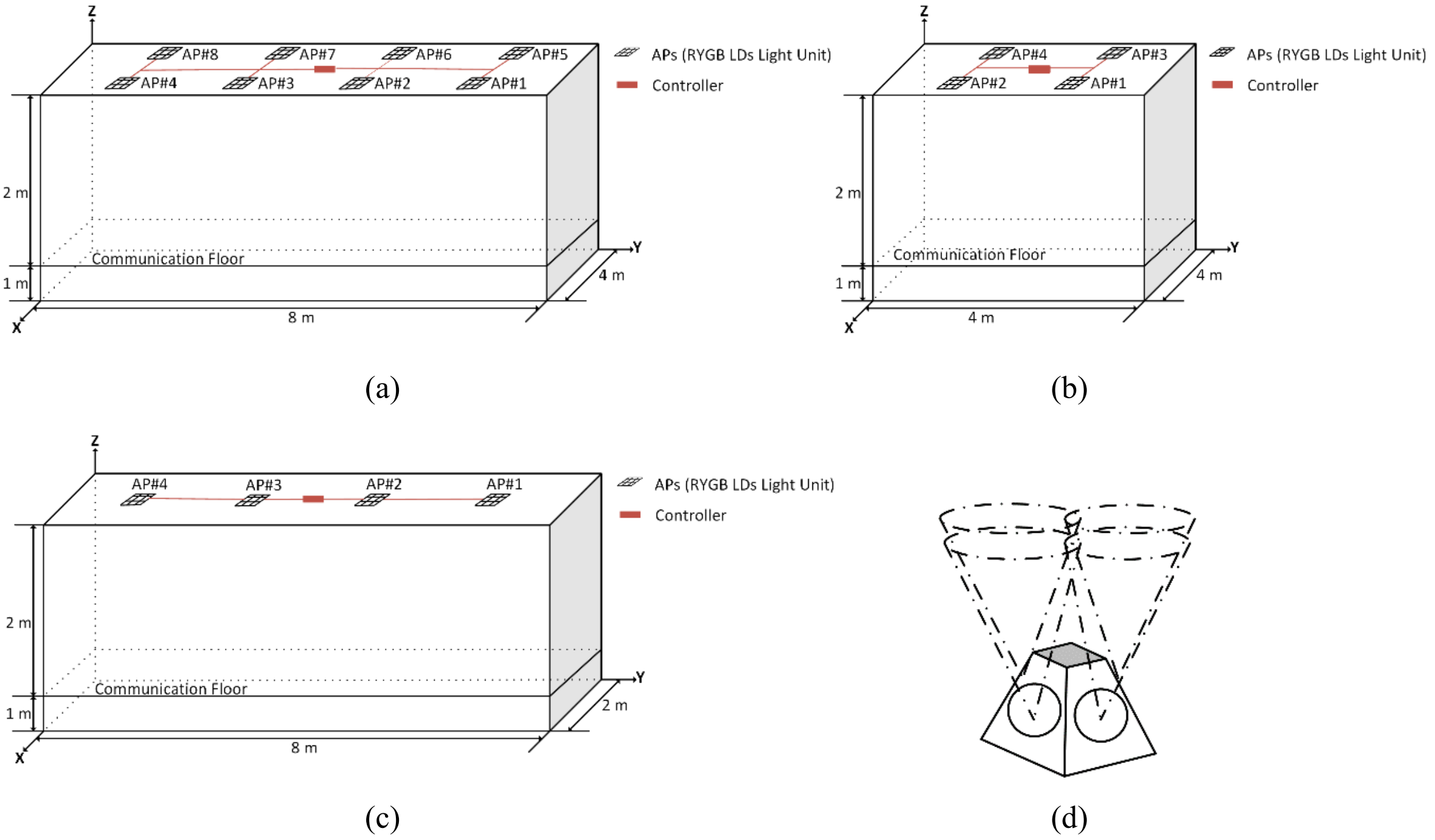

Figure 1: Configurations: (a) Room A, (b) Room B, (c) Room C and (d) Angle Diversity Receiver (ADR).

**Table 1.** System Parameters

| Parameters | Configurations | |
|---|---|---|
| Walls and ceiling reflection coefficient | 0.8 [36] | |
| Floor reflection coefficient | 0.3 [36] | |
| Number of reflections | 1 | 2 |
| Area of reflection element | 5 cm × 5 cm | 20 cm × 20 cm |
| Order of Lambertian pattern, walls, floor and ceiling | 1 [36] | |
| Semi-angle of reflection element at half power | 60° | |
| Number of RYGB LDs per unit | 12 | |
| Transmitted optical power of Red LD | 0.8 W | |
| Transmitted optical power of Yellow LD | 0.5 W | |
| Transmitted optical power of Green LD | 0.3 W | |
| Transmitted optical power of Blue LD | 0.3 W | |

| | | | | |
|---|---|---|---|---|
| Total transmitted power of RYGB LD | 1.9 W | | | |
| | **Room A** | | | |
| Width × Length × Height (x, y, z) | 4 m × 8 m × 3 m | | | |
| Number of transmitters' units | 8 | | | |
| Transmitters locations (x, y, z) | (1 m, 1 m, 3 m), (1 m, 3 m, 3 m), (1 m, 5 m, 3 m), (1 m, 7 m, 3 m), (3 m, 1 m, 3 m), (3 m, 3 m, 3 m), (3 m, 5 m, 3 m) and (3 m, 7 m, 3 m) | | | |
| | **Room B** | | | |
| Width × Length × Height (x, y, z) | 4 m × 4 m × 3 m | | | |
| Number of transmitters' units | 4 | | | |
| Transmitters locations (x, y, z) | (1 m, 1 m, 3 m), (1 m, 3 m, 3 m), (3 m, 1 m, 3 m) and (3 m, 3 m, 3 m) | | | |
| | **Room C** | | | |
| Width × Length × Height (x, y, z) | 2 m × 8 m × 3 m | | | |
| Number of transmitters' units | 4 | | | |
| Transmitters locations (x, y, z) | (1 m, 1 m, 3 m), (1 m, 3 m, 3 m), (1 m, 5 m, 3 m) and (1 m, 7 m, 3 m) | | | |
| | **Receiver** | | | |
| Responsivity Red | 0.4 A/W | | | |
| Responsivity Yellow | 0.35 A/W | | | |
| Responsivity Green | 0.3 A/W | | | |
| Responsivity Blue | 0.2 A/W | | | |
| Number of photodetectors | 4 | | | |
| Area of the photodetector | 20 $mm^2$ | | | |
| Photodetector | 1 | 2 | 3 | 4 |
| Azimuth angels | 45° | 135° | 225° | 315° |
| Elevation angels | 70° | 70° | 70° | 70° |
| Field of view (FOV) of each detector | 25° | | | |
| Receiver noise current spectral density | 4.47 pA/√Hz [34] | | | |
| Receiver bandwidth | 5 GHz | | | |

## 3. SIMULATION SETUP AND RESULTS

In this work, a wavelength division multiple access (WDMA) scheme is used to provide multiple access. Each AP uses laser diodes (LDs) that offer four wavelengths: Red, Yellow, Green and Blue (RYGB) which are mixed to provide white light as stated in [34]. Two different scenarios of 8 users were evaluated over the three different room sizes. In the first scenario, a set of 4 users were placed under an AP which is consider to be the worst scenario. In the second scenario, users were distributed over the room. A MILP model was utilised to optimise the resource allocation for each scenario in each room based on maximising the sum of all users' SINRs [37], [38]. Tables 2 and 3 show the locations of users and their resource allocation for both scenarios in different rooms. However, users' locations are given to the controller which is located on the ceiling of each room (See Figure 1a, b, c) [37]. Figure 2 illustrates an example of how the WDMA system works in VLC based on a scenario of three APs and users with two wavelengths (Red and Blue). Solid lines indicate the link that is carrying modulated data to a user from the assigned AP and wavelength, while dashed lines refer to an interference link that is carrying modulated data using the same wavelength from another AP. Dotted lines indicate background noise which is the unmodulated wavelength at an AP. In the example in Figure 2, User 1, who is the only one assigned to the blue wavelength, suffers only from the background noise, whereas, Users 2 and 3, who are assigned to the same wavelength (Red) from a different AP, suffer from the interference as well as background noise.

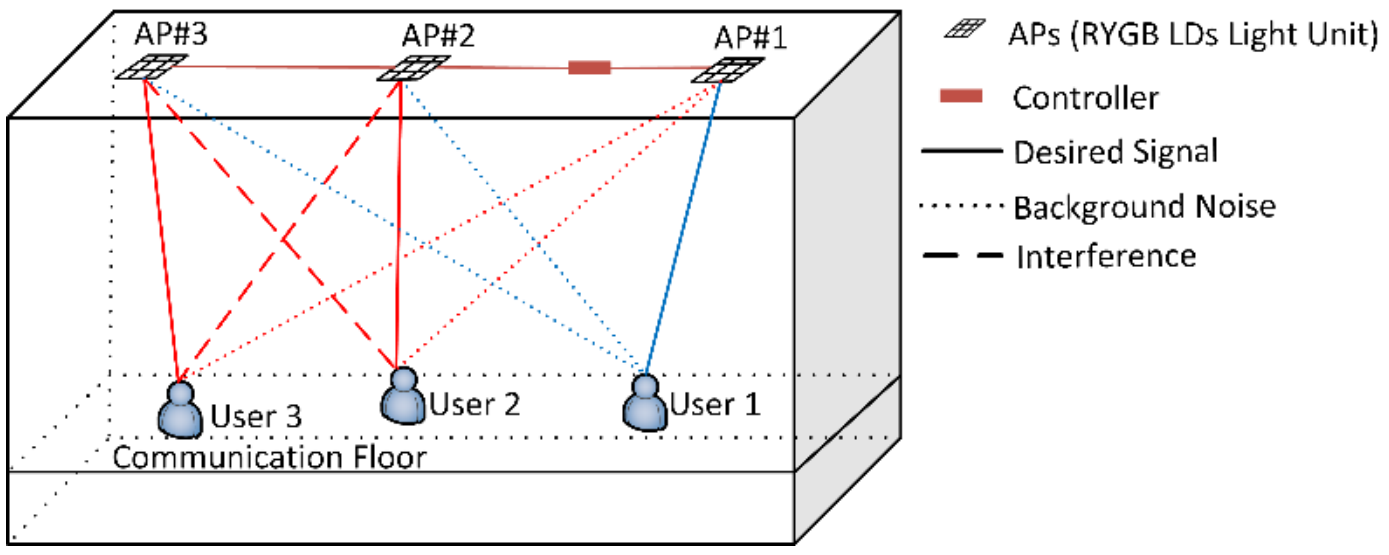

Figure 2: WDMA example.

**Table 2.** Scenario 1 with the optimised resource allocation

| User | Room A | | | | Room B | | | | Room C | | | |
|---|---|---|---|---|---|---|---|---|---|---|---|---|
| | Location (x, y, z) | AP | Branch | wavelength | Location (x, y, z) | AP | Branch | wavelength | Location (x, y, z) | AP | Branch | wavelength |
| 1 | (0.5,6.5,1) | 3 | 4 | Red | (0.5,2.5,1) | 1 | 2 | Red | (0.5,1.5,1) | 1 | 1 | Red |
| 2 | (0.5,7.5,1) | 4 | 4 | Yellow | (0.5,3.5,1) | 2 | 2 | Yellow | (0.5,0.5,1) | 2 | 1 | Yellow |
| 3 | (1.5,6.5,1) | 8 | 1 | Red | (1.5,2.5,1) | 2 | 1 | Red | (0.5,6.5,1) | 3 | 2 | Yellow |
| 4 | (1.5,7.5,1) | 4 | 3 | Red | (1.5,3.5,1) | 1 | 1 | Yellow | (0.5,7.5,1) | 4 | 2 | Red |
| 5 | (2.5,0.5,1) | 5 | 1 | Red | (2.5,0.5,1) | 3 | 3 | Red | (1.5,0.5,1) | 1 | 4 | Yellow |
| 6 | (2.5,1.5,1) | 1 | 3 | Red | (2.5,1.5,1) | 4 | 3 | Yellow | (1.5,1.5,1) | 2 | 4 | Red |
| 7 | (3.5,0.5,1) | 5 | 2 | Yellow | (3.5,0.5,1) | 3 | 4 | Yellow | (1.5,7.5,1) | 3 | 3 | Red |
| 8 | (3.5,1.5,1) | 6 | 2 | Red | (3.5,1.5,1) | 4 | 4 | Red | (1.5,6.5,1) | 4 | 3 | Yellow |

**Table 3.** Scenario 2 with the optimised resource allocation

| User | Room A | | | | Room B | | | | Room C | | | |
|---|---|---|---|---|---|---|---|---|---|---|---|---|
| | Location (x, y, z) | AP | Branch | wavelength | Location (x, y, z) | AP | Branch | wavelength | Location (x, y, z) | AP | Branch | wavelength |
| 1 | (0.5,1.5,1) | 1 | 4 | Red | (0.5,1.5,1) | 1 | 4 | Yellow | (0.5,0.5,1) | 1 | 1 | Yellow |
| 2 | (0.5,5.5,1) | 3 | 4 | Red | (0.5,2.5,1) | 2 | 3 | Red | (0.5,3.5,1) | 2 | 4 | Red |
| 3 | (0.5,6.5,1) | 4 | 1 | Red | (1.5,0.5,1) | 1 | 2 | Red | (0.5,6.5,1) | 4 | 3 | Red |
| 4 | (1.5,3.5,1) | 2 | 3 | Red | (1.5,3.5,1) | 2 | 4 | Yellow | (1.5,1.5,1) | 1 | 1 | Red |
| 5 | (2.5,1.5,1) | 5 | 4 | Red | (2.5,1.5,1) | 3 | 2 | Yellow | (1.5,2.5,1) | 2 | 4 | Yellow |
| 6 | (2.5,6.5,1) | 8 | 1 | Red | (2.5,2.5,1) | 4 | 4 | Yellow | (1.5,4.5,1) | 3 | 2 | Red |
| 7 | (3.5,3.5,1) | 6 | 3 | Red | (2.5,3.5,1) | 4 | 1 | Red | (1.5,5.5,1) | 3 | 3 | Yellow |
| 8 | (3.5,5.5,1) | 7 | 3 | Red | (3.5,0.5,1) | 3 | 4 | Red | (1.5,6.5,1) | 4 | 2 | Yellow |

In this work, the assignment of AP and wavelength to users in both scenarios in each room is optimised by using the MILP model. Subsequently, the optical channel bandwidth, the SINR and the supported data rates for each user in both scenarios in each room were determined. Figures 3 – 5 present the simulation results. Figure 3 shows the optical channel bandwidth for both scenarios in all rooms. The results show that when the room has large dimensions like Room A, the optical channel bandwidth is better than other narrower rooms. The reason behind that is that, when the room is larger and with the use of ADR, reflections cannot be detected from the walls in certain places. Room A provides a channel bandwidth for both scenarios, between 4.5 GHz – 8.5 GHz. Room C (corridor) has the lowest channel bandwidth, with bandwidth of around 4.2 GHz – 6.2 GHz.

Figure 4 presents the SINR for both scenarios in all rooms. Room A in Scenario 1, which is the worst scenario, provides SINR above 15.6 dB for all user locations. Rooms B and C which are smaller than Room A, have lower SINR for some user locations due to the high level of interference between users. For SINR lower than 15.6 dB a forward error correction (FEC) method was used to provide the same performance of 15.6 dB, similar to [38].

However, all rooms in Scenario 2 offer high SINR above 15.6 dB. All user locations in both scenarios in all rooms can support high data rates above 5.5 Gbps, as shown in Figure 5.

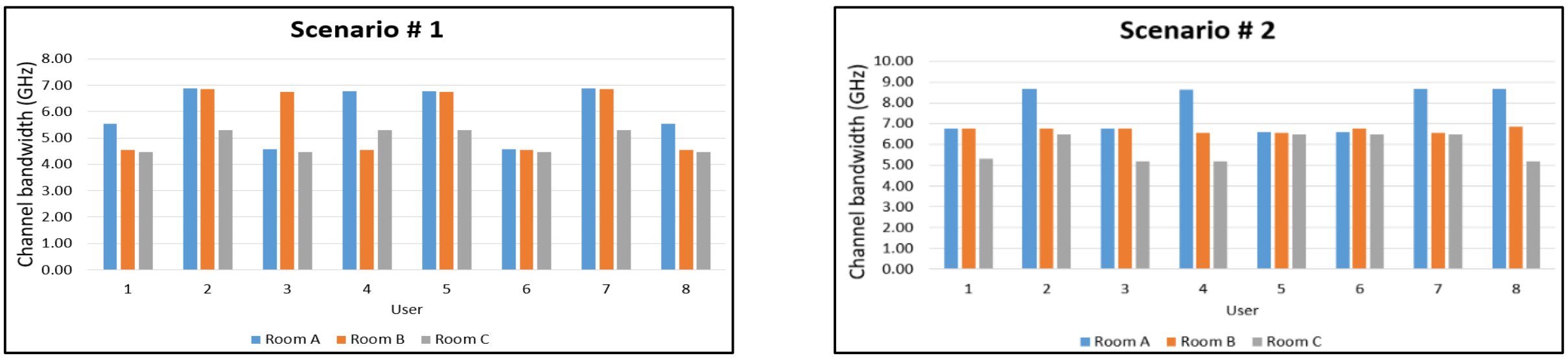

Figure 3: Channel bandwidth.

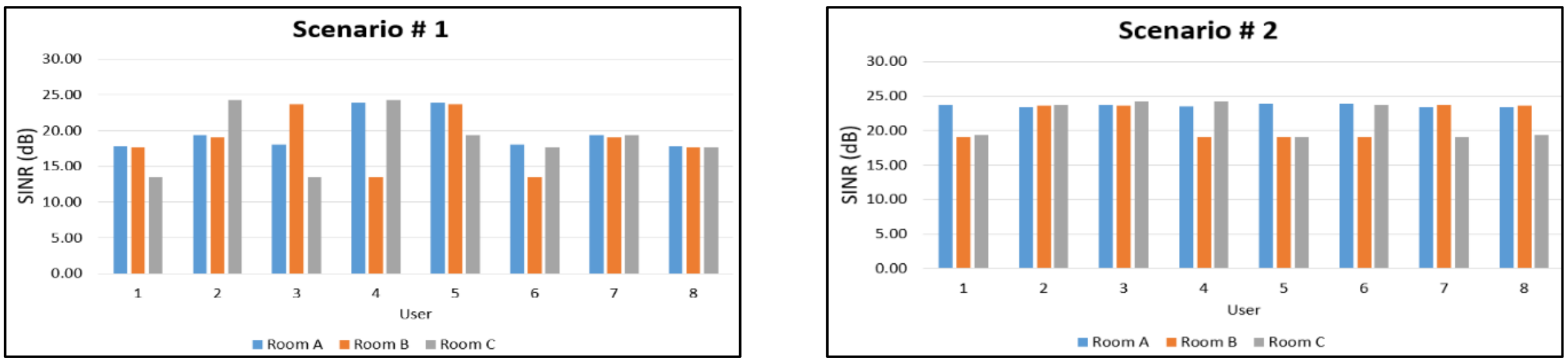

Figure 4: SINR.

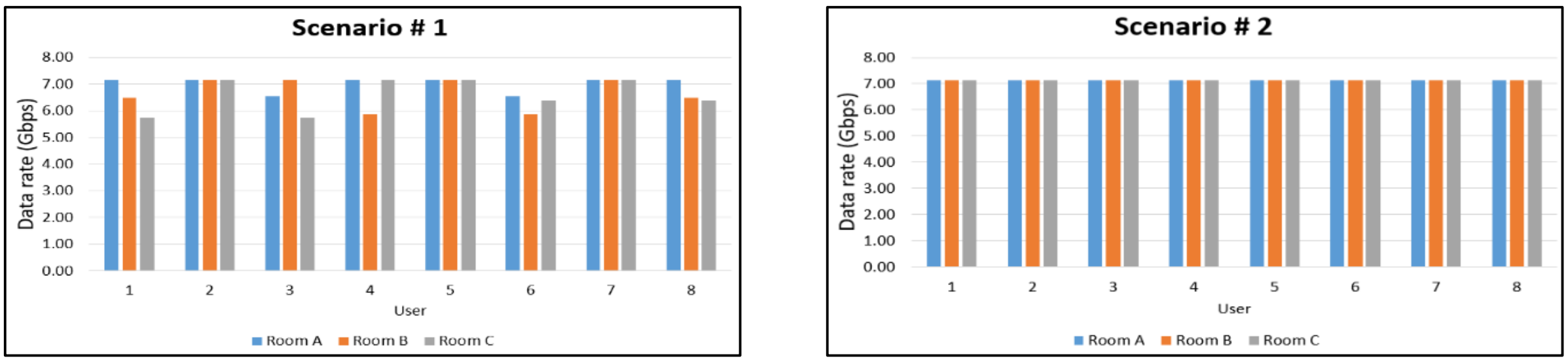

Figure 5: Data Rate.

## 4. CONCLUSIONS

This paper has investigated the impact of room size on wavelength division multiplexing (WDM) optical wireless systems. Three different room sizes were considered in this work, with two scenarios of 8 users in each room. Each AP can provide four wavelengths (red, yellow, green and blue) based on RYGB LDs that are used as transmitters. An optical receiver, an ADR, was used in this work. A mixed-integer linear programming (MILP) model was developed to optimise the resource allocation of wavelengths and access points to users. The optical channel bandwidth, SINR and data rate were measured for each user in both scenarios in all rooms. Room A provides a better channel bandwidth and SINR compared to the other rooms. However, all rooms offer a data rate above 5 Gbps in both scenarios.

## ACKNOWLEDGEMENTS

The authors would like to acknowledge funding from the Engineering and Physical Sciences Research Council (EPSRC) INTERNET (EP/H040536/1), STAR (EP/K016873/1) and TOWS (EP/S016570/1) projects. The authors extend their appreciation to the deanship of Scientific Research under the International Scientific Partnership Program ISPP at King Saud University, Kingdom of Saudi Arabia for funding this research work through ISPP#0093. MKAA would like to thank Taibah University in the Kingdom of Saudi Arabia for funding her PhD scholarship, OZA would like to thank Umm Al-Qura University in the Kingdom of Saudi Arabia for funding his PhD scholarship, SOMS would like to thank the University of Leeds and the Higher Education Ministry in Sudan for funding her PhD scholarship. SHM would like to thank EPSRC for providing her Doctoral Training Award scholarship. All data are provided in full in the results section of this paper.